# Sustainable financing of permanent $CO_2$ disposal through a Carbon Takeback Obligation

Stuart Jenkins*[1], Eli Mitchell-Larson[2,3], Stuart Haszeldine[4], and Myles Allen[1,2]

[1] – AOPP, Department of Physics, University of Oxford, Parks Road, Oxford, UK, OX1 3PU
[2] – ECI, School of Geography, University of Oxford, South Parks Road, Oxford, UK, OX1 3QY
[3] – Saïd Business School, University of Oxford, Park End St, Oxford, UK, OX1 1HP
[4] – School of Geosciences, University of Edinburgh, Edinburgh, UK, EH9 3FE
* - Corresponding author: stuart.jenkins@wadham.ox.ac.uk

**Abstract**

Unless there is immediate, unprecedented, reduction in global demand for carbon-intensive energy and products, then capture and permanent storage of billions of tonnes of carbon dioxide ($CO_2$) annually will be needed before mid-century to meet Paris Agreement goals. Yet competition from cheaper, temporary, carbon storage means that permanent disposal remains starved of investment, currently representing about 0.1% of Energy and Industrial Process (EIP) emissions. This stored fraction must reach 100% to stop EIPs causing global warming. Here we show that a cost-effective transition can occur by mandating an increasing stored fraction through a progressive Carbon Takeback Obligation (CTO) on fossil fuel producers and importers. Projected costs of storage to the consumer are lower than pricing carbon emissions in conventional 1.5°C scenarios until the 2040s, and comparable or lower thereafter. A CTO combined with measures to reduce $CO_2$ production would deliver the lowest-risk pathway to achieving net zero.

**Background: $CO_2$ storage in ambitious mitigation scenarios**

The IPCC's Special Report on a Global Warming of 1.5°C concluded: "Reaching and sustaining net-zero global anthropogenic $CO_2$ emissions *and* declining net non-$CO_2$ radiative forcing would halt anthropogenic global warming on multi-decadal time scales"[1]. Both conditions will need to be met around mid-century to meet the more ambitious goal of the Paris Climate Agreement, to "pursue efforts to limit the temperature increase to 1.5°C"[2–4]. All scenarios indicate that demand reduction and substitution alone will be insufficient. To limit cumulative emissions of $CO_2$ to deliver these ambitious goals requires some level of active compensatory $CO_2$ capture and storage, and eventually compensatory removal of excess $CO_2$ from the atmosphere[1]. It is apparent that even ambitious nations such as Sweden or the UK, are finding difficulty in following a 1.5-2°C pathway, with emissions trajectories substantially greater than required[5].

The need for immense tonnages of permanent geological carbon storage (GCS) to meet ambitious mitigation goals is well-documented[1]. Carbon removal is often conflated with one possible technology: Bio-energy with carbon capture and storage (BECCS)[6]. However, the disadvantages and hidden costs of reliance on BECCS are well-documented[7,8]. In this paper, we therefore assume that large-scale BECCS deployment is constrained, with initial $CO_2$ disposal accomplished through traditional CCS (point source capture with GCS), and subsequently through industrial Direct Air Capture (DAC) with GCS (sometimes referred to as Direct Air Capture with Carbon Storage, DACCS). We also assume that the scope for offsetting energy (in which we here include transport) and industrial process (EIP) emissions with "Nature-based Climate Solutions" (NCS) is limited, since of the order of 10 $GtCO_2$/year of $CO_2$ uptake, which is at the higher end of estimates of NCS potential[9], may be required simply to offset continued degradation of the biosphere and carbon release from earth system feedbacks due to warming[10]. Sustaining net zero global $CO_2$ emissions over a multi-decade period will require net zero or net negative EIP emissions. Recognising these facts simplifies the problem, since it allows us to focus exclusively on the challenge of achieving net zero in the EIP sector as a necessary condition for halting global warming.

Figure 1a shows net EIP emissions to the atmosphere in cost-effective global Integrated Assessment Model (IAM) scenarios that meet the Paris goals through the imposition of demand-side measures represented by a global carbon price[11,12]. Global EIP emissions in scenarios that limit warming to 1.5°C (dark blue) or well below 2°C (light blue) decline rapidly to reach net zero either by mid-century or before 2100 respectively. Red scenarios show an idealised 1.5°C-consistent supply-side policy, a global CTO, discussed below. Figure 1b shows that significant quantities of $CO_2$ continue to be produced from fossil sources even as emissions decline to net zero in the IAM scenarios, much of this in "hard-to-abate sectors", including aviation, marine shipping, steel and cement production, and load-balancing in power grids[13]. Net zero emissions to the atmosphere are achieved by progressively balancing this $CO_2$ production with GCS (figure 1c). Scenarios that explicitly exclude or minimise reliance on GCS[14,15] have shown it is possible to limit warming to 1.5°C, but only through some combination of unprecedented behaviour or technological innovation, leveraged at sensitive intervention points of the socio-economic system[16]. Neither of these can be objectively priced for direct comparison with least-cost scenarios, nor do they appear to be reliable prospects at present, suggesting a need for a "backstop" policy in case they fail to materialise[17].

Figure 1d suggests a form this backstop might take. Blue lines present the IAM scenarios, plotting the fraction of $CO_2$ generated by EIP that is geologically stored (the ratio 1c/1b[18]). This stored fraction has a very similar shape across almost all 1.5°C scenarios, rising parabolically to 100% by the date of net zero around mid-century[18]. This shape is inevitable for the most ambitious scenarios, because the stored fraction is currently close to zero and not increasing significantly, so the smoothest pathway to 100% (re)capture and storage, which is required by definition for net zero EIP emissions, is a quadratic increase. Some degree of smoothness is imposed by the constraints of capital and technology deployment rates, but our conclusions are robust to the precise shape of this curve: see supplementary information figure S1 for the corresponding figure assuming a cubic increase in stored fraction. The well-below-2°C scenarios show more degrees of freedom, but many also conform to this generic quadratic pathway. The red line shows a stylised scenario in which this fraction is increased quadratically through a supply-side policy we name a Carbon Takeback Obligation (CTO).

**Carbon Takeback: from "polluter pays" to "producer responsibility"**

To a good approximation, $CO_2$-induced warming is proportional to the total stock of carbon dioxide emitted into the atmosphere[19–22] minus that which is actively removed and returned to the geosphere. Demand-side policies such as carbon pricing are designed to reduce the flow of emissions, but are less well-suited to limit the total stock. The main reason is technology and policy uncertainty. In IAMs, GCS is deployed at scale, at a predictable cost and with presumed public consent, in anticipation of increasing global carbon prices: hence the relatively smooth increase in the stored fraction in the blue lines in figure 1d. But these IAMs are cost-optimised over the entire century. Deployment takes into account future carbon prices. There is no guarantee that this orderly deployment with perfect foresight will take place in reality. Figure 1d indicates that 11% of EIP $CO_2$ production should be geologically stored by 2030 on a smooth, quadratic, least-cost pathway to net zero by 2050. No country, even those with net zero targets, monitors or reports this stored fraction in their Nationally Determined Contribution (NDC), so, unsurprisingly, current planned sequestration deployment is completely inadequate. The UK, for example, will achieve a stored fraction of only 1-2% of EIP $CO_2$ emissions stored by 2030 because ambition is set by a policy ceiling of two pilot projects, capturing from industry[23]. By contrast independent statutory advisers to the UK government recommend storage of at least 10$MtCO_2$/yr operating by 2030, increasing to at least 75-175 Mt/yr $CO_2$ stored in 2050 required for net zero[24]. In a global perspective, $CO_2$ storage needs to be operating at a commercial scale by 2030 to have any prospect of reaching 5 $GtCO_2$/yr by 2050 and 10 $GtCO_2$/yr by 2100[25]. Storage needs are unequally distributed, operating in 2050 in developed economies and China at 2 $tCO_2$ per person per year plus an additional 1.7 tCO2 per year DAC, with storage of only 0.5 $tCO_2$ per person per year in emerging economies[26]. However in 2019, only 19 large CCS projects operated globally capturing and storing just 40 $MtCO_2$/yr, with 32 in construction or planning, making scale-up terminally slow[27]. The reason is that $CO_2$ capture and storage is still seen as a relatively expensive means of reducing the near-term flow of $CO_2$ emissions to the atmosphere, rather than as the only means available of addressing specific hard-to-abate sectors and permanently extending the overall stock of $CO_2$ production that is consistent with meeting any ambitious climate goal. All current projects depend to some degree on the tax-payer, through direct procurement, subsidies or tax-breaks. While such incentives are necessary for first-of-a-kind investments, a transition plan to a sustainable financing model must be built in from the start to maintain confidence on all sides and avoid perverse incentives and "subsidy lock-in".

Everyone involved in the production and emission of $CO_2$ from fossil sources is depleting the total remaining stock that can be emitted to the atmosphere before initiating dangerous climate change. Hence both producers and consumers have a collective responsibility to contribute to the development of GCS to ensure a smooth transition to net zero emissions as that stock is exhausted. Noting that one of the clearest beneficiaries of successful GCS development are the owners of fossil fuel assets, a simple way of discharging this responsibility is the Carbon Takeback Obligation (CTO). Under a CTO, primary extractors or importers of fossil carbon (including non-fuel sources such as lime for cement) are required to demonstrate permanent sequestration of a progressively increasing fraction of the $CO_2$ generated by the production, refining, transport and use of the products they sell. This fraction increases to 100%, ensuring net zero EIP $CO_2$ emissions, by a specified target date, and could exceed 100% if necessary to support net $CO_2$ removal thereafter. To be commensurate with the climate impact of $CO_2$ emissions, permanent storage must be interpreted as securing captured $CO_2$ in reservoirs with lifetimes greater than 10,000 years, a standard that is currently met only by GCS or alkaline metal remineralisation pathways[28]. If applied across an entire jurisdiction, the cost of a CTO would be shared between producers and consumers of carbon-intensive products.

To calculate a net-zero-compliant CTO pathway, it is necessary to know only the date of adoption, the date of net zero, and the shape of the profile. This minimises subjectivity, maximises transparency and makes it simple to explain, monitor and enforce. The red line in Figure 1d shows a quadratic increase from 0% in 2020 to 100% by 2050, consistent with the shape of the IAM cost-effective 1.5°C scenarios, reaching 11% by 2030 and 44% by 2040. In other words, the 1.5°C scenarios show a quadratically-increasing stored fraction as a hypothetical *result* of demand-side climate measures, a result which is by no means guaranteed and looking increasingly unlikely because policy uncertainty is discouraging investment in GCS. In contrast, the CTO pathway uses a similar pathway as a *driver* of behaviour to ensure that the desired outcome, net zero EIP emissions around 2050, is achieved. A more convex profile would back-load the burden of GCS deployment onto later decades, raising questions of temporal equity since those decades will also be experiencing the highest impacts of climate change over this transition period. Despite this concern, our conclusions are robust to the adoption of a cubic profile, which would correspond to a scenario in which a quadratic CTO is applied to a linearly increasing fraction of global fossil carbon use, only reaching 100% in 2050 (see Supp. Info. fig. S1).

A global CTO would be most simply implemented at the point of extraction, as a licensing condition for participation in international markets. But initially, a CTO could be introduced to a single country or trading bloc, levied and discharged most efficiently on those very few organisations owning fossil carbon at the point of extraction or import into that jurisdiction. If imposed on all extractors and importers of fossil fuels, costs of $CO_2$ disposal are initially small (because the stored fraction starts small) and can be passed downstream to consumers, limiting impact on competition and the threat of carbon leakage. Unlike carbon taxes and other levies aimed at changing consumer behaviour, however, there is a transparent link between a CTO and achieving net zero emissions. In the absence of a fossil fuel ban or prohibitively expensive tax-payer-funded $CO_2$ removal programme, a CTO is the only way of achieving net zero EIP emissions, the only way of stopping fossil fuels causing global warming, and hence the only way of any jurisdiction permanently "ending its contribution to global warming"[29].

This type of obligation, a form of Extended Producer Responsibility, has precedent, for example with the Waste Electrical and Electronic Equipment (WEEE) Directive in the EU[30], and the California Low-Carbon Fuel Standard (LCFS) in the US[31]. A CTO may be thought of as requiring vendors of fossil carbon to dispose of the unwanted $CO_2$ "packaging" that accompanies the energy they sell, or applying a LCFS upstream to decarbonise fossil fuels themselves. Crucially, under a CTO producers would take responsibility not just for their direct emissions from owned, controlled or purchased sources (Scope 1 & 2, already covered in some jurisdictions by conventional demand-side carbon pricing), but also for indirect emissions generated by the products they sell (Scope 3)[32]. This is consistent with the ambitions of some European fossil fuel companies that have begun to set aspirational net zero targets that encompass scope 3 emissions[33]. A regulatory CTO framework would create a level playing-field, allowing companies to invest in GCS without the concern that this might put them at a competitive disadvantage.

**Implications of a Carbon Takeback Obligation**

A crucial feature of the CTO is that it does not, itself, prescribe or proscribe the rate of production of $CO_2$, only the fraction of that production that is stored. It is therefore a type of intensity target, which are generally more acceptable to developing countries than absolute emission limits, improving the prospects for wider participation in a CTO than in any global cap-and-trade regime. This flexibility has other implications. Red shaded plumes in figure 1 show Paris-compliant emissions pathways in which the stored fraction follows the same quadratic path to 100%, or net zero EIP emissions, in 2050. By contrast $CO_2$ production is bounded by two scenarios: one which assumes that consumption declines as the stored fraction rises (solid bound) to give a linear decline in net EIP annual emissions from 2020 to 2050; and another where gross EIP $CO_2$ production continues along the median SSPX-45 pathway (dashed bound) despite very high expenditure on GCS (we are not suggesting this is a plausible scenario, but it illustrates the range of pathways open). After 2050, the stored fraction either remains at 100% (solid) or rises above 100% (dashed) to a level that results in the same net cumulative emissions to 2100. Both scenarios deliver net zero EIP emissions in 2050 and the same post-2100 climate outcome, but the dashed, high-consumption scenario results in higher peak warming and requires substantially more $CO_2$ storage.

Panels e and f compare the estimated costs of a global CTO with the cost of abatement in conventional price-driven scenarios. Panel e compares the cost of complying with a CTO per ton of $CO_2$ produced (red) with the effective global carbon price in conventional scenarios (blue). This per-tonne CTO compliance cost is a function of the stored fraction (1d) and the full-chain cost to capture, transport, and permanently store one tonne of $CO_2$. If the initial objective of a CTO is incentivising the development of a $CO_2$ storage industry to complement other climate change mitigation measures, then any $CO_2$ that would otherwise have been emitted to the atmosphere could be used to generate a tradeable storage certificate to discharge a CTO, with safeguards to ensure $CO_2$ is not generated purely for the purpose of storage[34]. Hence we assume an initial full chain cost range of $C_1$ = \$40-60/tCO2 stored, representing the cheapest, high-purity $CO_2$ capture opportunities[35–37].

By 2050, we assume conservatively that all remaining fossil fuel use and industrial process emissions result from non-stationary or diffuse applications like aviation for which capture at source is not possible, giving a cost $C_2$ = \$200-600/tCO$_2$ stored which reflects a range of plausible costs for DAC+GCS[37,38]. Although all costs are likely to drop over time, the dominant effect will be a transition toward ever more diffuse streams of $CO_2$ once all remaining point sources have been captured, culminating in reliance on DAC. We model this transition as a simple straight line progression from $C_1$ and $C_2$ as the stored fraction S increases, giving a stylised cost of CTO compliance $C(S) = S(C_1+(C_2-C_1)S)$ per tonne of $CO_2$ produced. At 11% sequestration, for example, the CTO compliance cost for a whole jurisdiction represents only 11% of the capture and storage cost of $CO_2$, or \$6-13 per tonne of $CO_2$ generated under these assumptions, because it is distributed over all fossil fuels used in a jurisdiction, not just those that are subject to capture. Despite these conservatively high cost assumptions, this supply-side CTO compliance cost is much lower than the demand side carbon prices required to deliver the equivalent climate outcome until the 2040s, and remains comparable or lower thereafter (see figure 1e).

Conventional mitigation scenarios focused on pricing emissions consistently show carbon prices of about \$1000/tCO$_2$ are required to achieve net zero emissions around mid-century (see figure S2 in supplementary information). Such prices could still occur in the absence of a CTO even if DAC+GCS costs are limited to \$200-600/tCO$_2$ because DAC and particularly GCS take time to deploy at scale. If, as at present, deployment rates fail to anticipate future carbon prices, then GCS and DAC are only deployed when all cheaper methods of reducing emissions are exhausted, and hence may not be available at scale when needed, leading to a need for very high carbon prices to "drive out" the last emissions[39]. Introducing a CTO now eliminates the risk of carbon prices ever exceeding the cost of DAC+GCS.

Figure 1f shows the total cost of the CTO compared to modelled reductions in global income in ambitious mitigation scenarios compared to the relevant baseline SSPX-45 scenario. This is a heuristic comparison, because we have not modelled the interaction of the CTO with other climate policies and the rest of the economy (although any such interaction would be heavily contingent on the details of how a CTO were implemented). It does, however, serve to illustrate two important points. First, the solid red curve shows that, if accompanied by a halving of total EIP $CO_2$ production over the next 30 years (see figure 1b), the cost of compliance with a CTO (peaking at around 1.5% of global GDP by mid-century) could be lower than even the most optimistic estimates of the cost of achieving the same climate goals in a conventional scenario driven by a global carbon price. This does not mean that the total economic impact of a CTO would be less than that of a global carbon price, because this compliance cost alone does not capture it. It does, however, indicate that a "second best" regulatory measure may be close in cost to an optimal pathway while also offering non-monetary advantages, such as certainty for investment in $CO_2$ storage and a transparent link between policy and outcome. These advantages may outweigh any economic inefficiency.

The second point is that it would be myopic to rely on a CTO alone. Compliance with a CTO would have very little impact on consumption in the short term because the cost per tonne of $CO_2$ generated ($6-13 for 11% sequestration in 2030) is lower even than current carbon prices. However, if consumption remains at SSPX-45 levels (dashed red curve), the total amount spent on CTO compliance rises rapidly. With the pessimistic high price final DAC+GCS cost of $600/$tCO_2$ and (implausibly) no impact on $CO_2$ production, the cost of CTO compliance could reach 10% of global GDP in 2050, approaching (but not exceeding) global income losses due to conventional emissions price- and tax-based mitigation measures in the highest-cost 1.5°C scenarios.

The principle of progressively increasing the stored fraction of $CO_2$ produced by burning fossil fuels through some form of producer obligation was first put forward in 2009[40] and expanded towards a CTO in 2015, when it was even proposed in draft UK legislation[41], although concerns were raised at the time about the potential cost and implications for competitiveness. CTO remains one policy option formally under consideration by UK government, and is still to be explored[42]. We have shown that the cost of compliance with a CTO can be comparable to and in most cases less than the cost of achieving similar climate objectives through traditional demand-side carbon pricing alone. That said, exclusive reliance on a CTO could result in increased near-term fossil fuel extraction and lock in high-carbon infrastructure resulting in higher peak warming and higher policy cost overall. All continued fossil fuel use without a proven and scalable $CO_2$ storage capability contributes to the risk of a precipitous transition when the stock of $CO_2$ that can be dumped into the atmosphere is exhausted. The unique value of the CTO is to create a market for $CO_2$ storage, spreading the cost of reducing this transition risk over all current fossil fuel users through a simple regulatory framework and at no cost to the taxpayer. It can therefore be implemented alongside existing actions such as the EU-ETS, provincial taxes, or federal carbon pricing. A CTO, combined with such measures to reduce $CO_2$ production, would deliver the lowest-risk pathway to achieving net zero.

## Methods

The IIASA SSP database[9] contains a number of variants of the SSPs as outputted by 5 IAMs (AIM/CGE, GCAM4, IMAGE, MESSAGE-GLOBIOM, REMIND-MAGPIE, WITCH-GLOBIOM), representing complementary narratives for societal evolution over the 21st century whilst sampling climate policy of varying ambition[8,10]. Figure 1 shows cost-effective scenarios driven by a global carbon price, not including scenarios that explicitly minimise use of GCS[14,15] and deliver climate goals through exogenously imposed additional measures such as consumer behaviour change, noting that "it is nearly impossible to put a price tag on most of these measures, none of the scenarios [with these additional measures] has been evaluated in terms of costs"[15]. $CO_2$ production in 1b excludes $CO_2$ generated by bioenergy (hence assuming this is captured contemporaneously — scenarios include fossil $CO_2$ associated with bioenergy production and transport), while $CO_2$ sequestration in 1c includes $CO_2$ stored from bioenergy, providing the net total rate of GCS. Stored fraction in 1d is simply the ratio of 1c/1b[18]. Carbon prices in 1e are taken directly from the SSP database, while the reduction in global GDP in 1f is taken relative to the corresponding SSPX-45 scenario. We are using SSPX-45 as a "current policies" scenario under which emissions peak around 2030, remain stable to mid-century and then decline but fail to reach zero by 2100, leading to a 2100 warming around 3°C and continued warming thereafter. Hence we are comparing the impact of a CTO assumed to be additional to current policies with the impact of more stringent climate policies required to meet 1.5°C and below-2°C climate goals. Figure S1 is identical to figure 1 but assumes a cubic increase in global stored fraction under the CTO scenario. Figure S2 shows that an approximately straight-line relationship between 2050 carbon prices and 2050 EIP emissions emerges as climate policy ambition is varied under each of the SSP socio-economic scenarios, with the SSPs ranked by global primary energy consumption over the century. Scenarios with higher primary energy consumption show higher emissions for carbon prices in the range of $10-100 per $tCO_2$, but also greater sensitivity of 2050 emissions to 2050 carbon prices (the slopes of the best-fit lines). Across all SSPs, achieving net zero emissions in 2050 appears to require a carbon price around $1,000 per $tCO_2$ under scenarios that rely on demand-side carbon pricing alone. By contrast, using $CO_2$ storage on the supply-side limits prices to less than the cost of DAC+GCS, estimated at $200-600 per $tCO_2$

## Acknowledgements

Shared Socioeconomic Pathways are curated by www.iiasa.ac.at. H is funded by EPSRC EP/P026214/1; A & J are funded by the UK NERC (NE/P019900/1) and EC H2020 (4C, 821003, and NEGEM, 869192). M-L is funded by the Pershing Square Foundation. The authors thank the contributions of Richard Millar in the initial analysis of IAM output and contributions to the design of the CTO.

## References

1. IPCC. Summary for Policymakers of the Special Report on the Global Warming of 1.5°C. *IPCC* (2018).
2. Haustein, K. *et al.* A real-time Global Warming Index. *Sci. Rep.* **7**, 15417 (2017).
3. Leach, N. J. *et al.* Current level and rate of warming determine emissions budgets under ambitious mitigation. *Nat. Geosci.* **11**, 574 (2018).
4. WMO. *WMO Provisional Statement on the State of the Global Climate in 2019*. (2019).

5. Anderson, K., Broderick, J. F. & Stoddard, I. A factor of two: how the mitigation plans of 'climate progressive' nations fall far short of Paris-compliant pathways. *Clim. Policy* **0**, 1–15 (2020).
6. Creutzig, F. *et al.* Bioenergy and climate change mitigation: an assessment. *Glob. Change Biol. Bioenergy* **7**, 916–944 (2015).
7. Smith, P. *et al.* Biophysical and economic limits to negative $CO_2$ emissions. *Nat. Clim. Change* **6**, 42–50 (2016).
8. Anderson, K. & Peters, G. The trouble with negative emissions. *Science* **354**, 182–183 (2016).
9. Griscom, B. W. *et al.* Natural climate solutions. *Proc. Natl. Acad. Sci.* **114**, 11645–11650 (2017).
10. Lowe, J. A. & Bernie, D. The impact of Earth system feedbacks on carbon budgets and climate response. *Philos. Trans. R. Soc. Math. Phys. Eng. Sci.* **376**, 20170263 (2018).
11. Riahi, K. *et al.* The Shared Socioeconomic Pathways and their energy, land use, and greenhouse gas emissions implications: An overview. *Glob. Environ. Change* **42**, 153–168 (2017).
12. Huppmann, D. *et al. IAMC 1.5°C Scenario Explorer and Data hosted by IIASA*. (Integrated Assessment Modeling Consortium & International Institute for Applied Systems Analysis, 2018). doi:10.22022/SR15/08-2018.15429.
13. Davis, S. J. *et al.* Net-zero emissions energy systems. *Science* **360**, (2018).
14. Grubler, A. *et al.* A low energy demand scenario for meeting the 1.5 °C target and sustainable development goals without negative emission technologies. *Nat. Energy* **3**, 515–527 (2018).
15. van Vuuren, D. P. *et al.* Alternative pathways to the 1.5 °C target reduce the need for negative emission technologies. *Nat. Clim. Change* **8**, 391–397 (2018).
16. Farmer, J. D. *et al.* Sensitive intervention points in the post-carbon transition. *Science* **364**, 132–134 (2019).
17. Asheim, G. B. *et al.* The case for a supply-side climate treaty. *Science* **365**, 325–327 (2019).
18. Millar, R. J. & Allen, M. R. Understanding the role of CCS deployment in meeting ambitious climate goals (Chapter 2). in *Carbon Capture and Storage* (Royal Society of Chemistry, 2019).
19. Stocker, T.F. AR5 Climate Change 2013: The Physical Science Basis — IPCC. *Camb. Univ. Press Camb. U. K. N. Y. NY USA* 1535 (2013).
20. Allen, M. R. *et al.* Warming caused by cumulative carbon emissions towards the trillionth tonne. *Nature* **458**, 1163–1166 (2009).
21. Matthews, H. D., Gillett, N. P., Stott, P. A. & Zickfeld, K. The proportionality of global warming to cumulative carbon emissions. *Nature* **459**, 829–832 (2009).
22. Millar, R. J. & Friedlingstein, P. The utility of the historical record for assessing the transient climate response to cumulative emissions. *Philos. Trans. R. Soc. Math. Phys. Eng. Sci.* **376**, 20160449 (2018).
23. CCUS Cost Challenge Taskforce. *GOV.UK* https://www.gov.uk/government/groups/ccus-cost-challenge-taskforce.
24. The CCC. Net Zero - The UK's contribution to stopping global warming. *Committee on Climate Change* https://www.theccc.org.uk/publication/net-zero-the-uks-contribution-to-stopping-global-warming/.
25. Huppmann, D., Rogelj, J., Kriegler, E., Krey, V. & Riahi, K. A new scenario resource for integrated 1.5 °C research. *Nat. Clim. Change* 1 (2018) doi:10.1038/s41558-018-0317-4.

26. Pye, D., Price, J., Cronin, J., Butnar, I. & Welsby, D. *Modelling 'leadership-driven' scenarios of the global mitigation effort*. https://www.theccc.org.uk/publication/modelling-leadership-driven-scenarios-of-the-global-mitigation-effort-ucl-energy-institute/ (2019).
27. Global CCS Institute. *Global Status of CCS 2019*. (2019).
28. Renforth, P. The potential of enhanced weathering in the UK. *Int. J. Greenh. Gas Control* **10**, 229–243 (2012).
29. UK Govt. UK becomes first major economy to pass net zero emissions law. *GOV.UK* https://www.gov.uk/government/news/uk-becomes-first-major-economy-to-pass-net-zero-emissions-law (2019).
30. European Commission. Waste electrical and electronic equipment (WEEE) directive. https://ec.europa.eu/environment/waste/weee/index_en.htm (2017).
31. Townsend, A. & Havercroft, I. THE LCFS AND CCS PROTOCOL: AN OVERVIEW FOR POLICYMAKERS AND PROJECT DEVELOPERS. *Glob. CCS Inst.* 24 (2019).
32. Greenhouse Gas Protocol. https://ghgprotocol.org/ (2020).
33. Dietz, S., Gardiner, D., Jahn, V. & Noels, J. Carbon Performance of European Integrated Oil and Gas Companies: Briefing paper. *Transit. Pathw. Initiat.* (2020).
34. SCHNEIDER, L. R. Perverse incentives under the CDM: an evaluation of HFC-23 destruction projects. *Clim. Policy* **11**, 851–864 (2011).
35. Irlam, L. Global Costs of Carbon Capture and Storage. *Glob. CCS Inst.* (2017).
36. Leeson, D., Mac Dowell, N., Shah, N., Petit, C. & Fennell, P. S. A Techno-economic analysis and systematic review of carbon capture and storage (CCS) applied to the iron and steel, cement, oil refining and pulp and paper industries, as well as other high purity sources. *Int. J. Greenh. Gas Control* **61**, 71–84 (2017).
37. Bui, M. *et al.* Carbon capture and storage (CCS): the way forward. *Energy Environ. Sci.* **11**, 1062–1176 (2018).
38. Fasihi, M., Efimova, O. & Breyer, C. Techno-economic assessment of CO2 direct air capture plants. *J. Clean. Prod.* **224**, 957–980 (2019).
39. Kriegler, E., Edenhofer, O., Reuster, L., Luderer, G. & Klein, D. Is atmospheric carbon dioxide removal a game changer for climate change mitigation? *Clim. Change* **118**, 45–57 (2013).
40. Allen, M. R., Frame, D. J. & Mason, C. F. The case for mandatory sequestration. *Nat. Geosci.* **2**, 813–814 (2009).
41. Haszeldine, R. S., Allen, M., Hepburn, C., Le Quéré, C. & Millar, R. *Certificates for CCS at reduced public cost: securing the UK's energy and climate future, Energy Bill 2015*. https://era.ed.ac.uk/handle/1842/15698 (2015).
42. Vivid Economics. *Greenhouse Gas Removal policy options*. https://www.vivideconomics.com/casestudy/greenhouse-gas-removal-policy-options/ (2019).

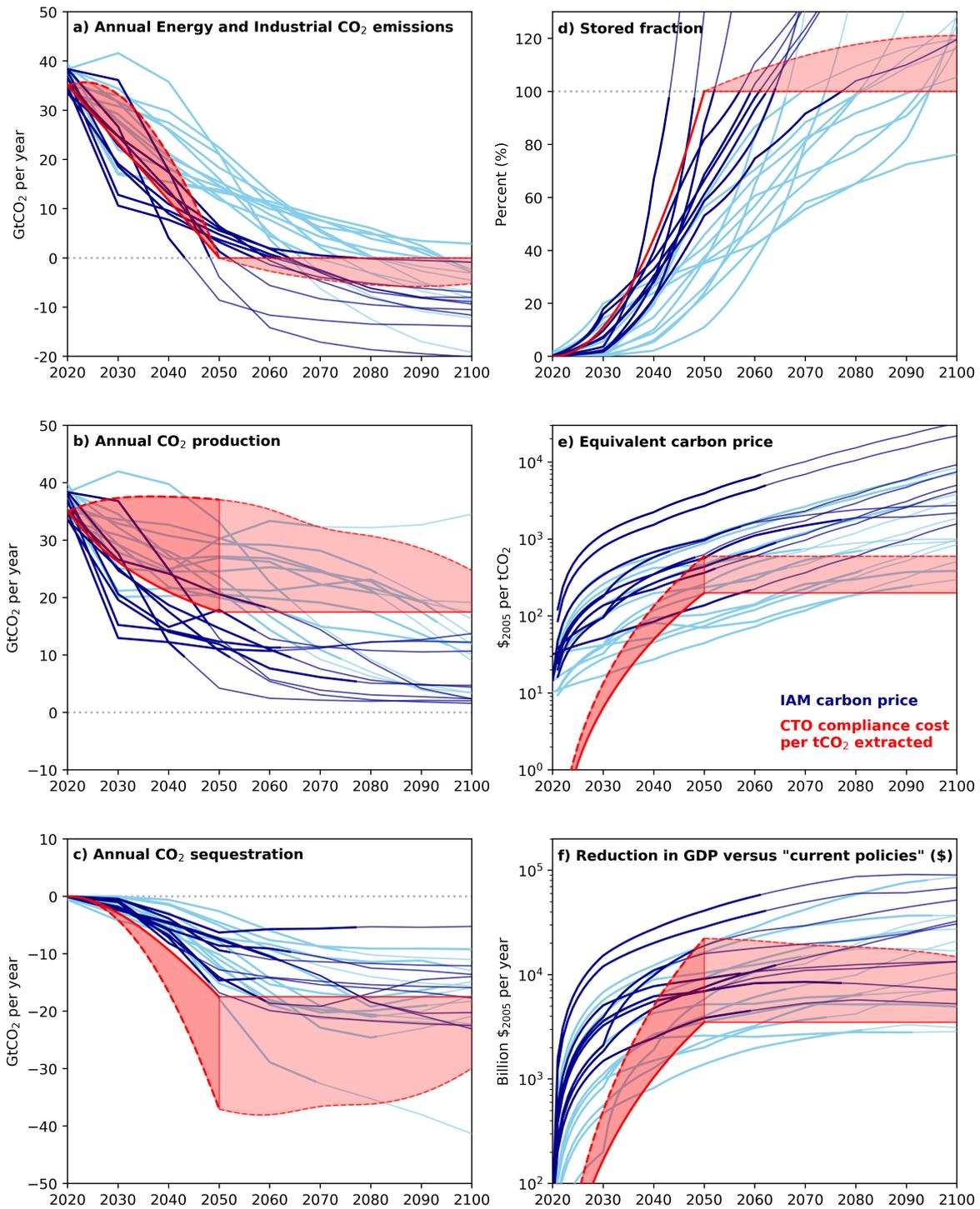

**Figure 1:** Comparing conventional mitigation scenarios with a global Carbon Takeback Obligation (CTO). Blue lines show cost-effective IAM scenarios plotted for multiple models and Shared Socioeconomic Pathways (SSPs) and two levels of climate ambition: SSPX-1.9 "1.5°C" (dark blue) and SSPX-2.6 "well-below 2°C" (light blue). A stylised CTO policy is overlaid in red. Panel a plots the global annual Energy and Industrial Processes (EIP) $CO_2$ emissions between 2020-2100 ($GtCO_2$/yr), which is the difference between gross annual EIP $CO_2$ production from fossil sources (b) and annual $CO_2$ sequestration (c)., which also includes sequestration from bioenergy. Panel d plots the stored fraction, or fraction of EIP $CO_2$ produced from fossil sources that is stored using GCS in each year. Panel e plots the carbon price driving the mitigation in the IAMs shown in panel a ($\$_{2005}$), compared with the cost of compliance with the CTO, while panel f shows the reduction in GDP relative to the SSPX-4.5 used as a proxy for "current policies" scenario compared with aggregate expenditure on compliance with the CTO. In all panels the solid line scenario assumes that consumption declines as the stored fraction rises (solid bound) to give a linear decline in net EIP annual emissions from 2020 to 2050; whilst the dashed red line scenario assumes gross EIP $CO_2$ production continues along the median SSPX-45 pathway (dashed bound) despite very high expenditure on GCS. The shaded region encompasses the plausible range of 1.5°C-consistent pathways in between these two bounding cases.